# Production of $^{230}$Pa as a Source for Medical Radionuclides $^{230}$U and $^{226}$Th Including Isolation by Liquid-liquid Extraction

A.N. Vasiliev [1,2], S.V. Ermolaev [1], E. V. Lapshina [1], M. G. Bravo [2], A. K. Skasyrskaya [1]

[1] *Institute for Nuclear Research of Russian Academy of Sciences, 60th October Anniversary Prospect, 7a, Moscow, 117312, Russia, vasiliev@inr.ru*

[2] *Lomonosov Moscow State University, GSP-1, Leninskie Gory, Moscow, 119991,* Russia

**Abstract**

The experimental cross sections of $^{230}$Pa formation in the reaction $^{232}$Th (p,3n) $^{230}$Pa for the protons energy range of 140-35 MeV are determined and compared with literature data. The yield of $^{230}$U and impurities of $^{232}$U and $^{233}$U depending on the energy of protons entering a thick thorium target are estimated. Liquid-liquid extraction with 1-octanol solution and addition of $Al^{3+}$ as a masking agent followed by back-extraction with 7M $HNO_3$+0.01M HF solution allows selective Pa isolation from a solution of irradiated Th and proposes simultaneous production of medical $^{225}$Ac and $^{230}$Pa/$^{230}$U from the same Th-target.

**Introduction**

α-Particles exhibit high values of linear energy transfer in a very limited range of action (~ 10 cell diameters), and this property promotes the development of targeted alpha therapy (TAT) – a fast-paced branch of nuclear medicine. The choice of a radionuclide for TAT is determined by its nuclear characteristic and chemical properties. The important aspects are the half-life of the radionuclide and the way of its decay as well as the availability of up-scaling production [1].

$^{230}$U is one of potential medical α-emitters. The decay of $^{230}$U generates a chain of short-lived products and it is accompanied by the emission of five α-particles with a total energy of 33.5 MeV, resulting in effective cell damage [2]. $^{230}$U can be used as an independent therapeutic radionuclide or as a source for a $^{230}$U/$^{226}$Th generator. The decay of the short-lived $^{226}$Th ($T_{1/2}$ = 30.6 min) generates a rapid cascade of four α-particles with a total energy of 27.7 MeV, for comparison, the decay of $^{213}$Bi ($T_{1/2}$ = 45.6 min) produces only one α-particle with an energy of 8.4 MeV. $^{226}$Th due to a relatively short half-life is suggested to be delivered by rapidly diffusing

peptide vectors as carrier molecules. Such therapy is promising for the treatment of easily accessible tumors.

The most productive way of direct formation of $^{230}$U consists in irradiation of $^{231}$Pa ($T_{1/2}$ = 3.3·10$^4$ y) with accelerated protons and deuterons by reactions: $^{231}$Pa(p,2n)$^{230}$U and $^{231}$Pa(d,3n)$^{230}$U. The initial $^{231}$Pa is a member of the $^{235}$U decay chain, it must be isolated from aged uranium samples, which is a disadvantage of the method, taking into account the limited availability of the raw materials.

Another approach is reactions of thorium nuclei with accelerated protons and deuterons resulting in the production of a precursor of $^{230}$Pa, decaying in $^{230}$U with a probability of 7.8%: $^{232}$Th (p,3n) $^{230}$Pa→$^{230}$U and $^{232}$Th (d,4n) $^{230}$Pa→$^{230}$U. The maximum amount of $^{230}$U accumulates in 27 days after the end of short-run irradiation. Many leading scientific organizations around the world such as TriLab cooperation [3-6], ARRONAX (France) [7], TRIUMF (Canada), are actively working under the development of this method. The authors from ARRONAX [7] have compared the yield of $^{230}$U from thick targets of $^{231}$Pa and $^{232}$Th. It was reported that proton irradiation leads to close yields of about 0.24 MBq/(µA·h), which is higher than the corresponding yields for the reactions with deuterons. The maximum of the $^{232}$Th(p,3n)$^{230}$Pa reaction excitation function corresponds to the proton energy of 19.9±0.3 MeV [2], therefore, industrial production of $^{230}$U can be organized on productive commercial cyclotrons operating at currents up to 1200 µA of protons with energies of 30 MeV, such as Cyclone 30 manufactured by IBA (Belgium) [8,9].

With an increase in proton energy, it becomes possible to obtain not only $^{230}$U but also other promising alpha emitters $^{225}$Ac [10] and $^{223}$Ra [11]. Radiopharmaceuticals containing $^{225}$Ac and its daughter $^{213}$Bi are successfully passing clinical trials for the treatment of leukemia, NHL, carcinoma, neuroendocrine tumor, glioma melanoma, and other malignant diseases [12,13]. Radiopharmaceutical Xofigo on the base of $^{223}$Ra is already approved for a common application in many countries for bone metastasis therapy [14]. The maximum yield of $^{225}$Ac is achieved at higher proton energy (more than 80 MeV [15]). This proton energy range is not optimal for the production of $^{230}$Pa, and $^{230}$Pa/$^{230}$U can be considered as an important by-product. The approach is considered promising in scientific centers where accelerators with a high flux of medium-energy protons are available (BNL, LANL, TRIUMF, INR RAS, etc.).

The chemical properties of protactinium are more complex and exhibit greater unpredictability than the properties of most elements. Protactinium is easily hydrolyzed in the absence of complexing agents, forming colloid and polymeric particles [16]. It is adsorbed on almost any available surface, which often causes the loss of Pa during chemical manipulations.

Various chemical methods are implemented for Pa separation: co-precipitation, liquid-liquid extraction (LLE), ion exchange, and extraction chromatography. The works published in literature on the isolation of $^{230}$Pa from irradiated thorium use primarily chromatographic methods. The authors suggest using anion-exchange sorbents AG MP-1 [17], AG 1x8 [4], commercially available extraction-chromatographic sorbents TRU resin [4], UTEVA resin [4], TK 400 [18], Cl resin [19], DGA resin and its analogs [6].

Chromatographic methods are not relevant from a technological point of view in the case of massive targets resulting in a large volume of concentrated initial solutions. Extraction methods are universal for protactinium isolation both from natural materials and from irradiated targets since they can be used at all stages of concentration and during final purification. The chemistry of Pa is relatively poorly investigated, although there is a large bulk of data in the literature on the LLE of Pa by various agents, for example, aliphatic alcohols, ketones, quaternary ammonium compounds, organophosphorus compounds [20].

A method for producing $^{230}$Pa/$^{230}$U simultaneously with $^{225}$Ac allowing to produce therapeutic activity of these radionuclides in one irradiation run is considered in the present paper. The objectives of the work are to determine the possible yield of $^{230}$Pa/$^{230}$U in the medium-energy proton range and to develop a fast and technological procedure for Pa isolation as a by-product of $^{225}$Ac.

**Experimental**

All chemicals were of p. a. quality or higher, obtained from Merck (Darmstadt, Germany), and used without additional purifications. All experiments were carried out using de-ionized "Milli-Q" water (18 M$\Omega\cdot$ cm$^{-1}$).

*Cross section measurement*

The metal foils with a thickness of 0.045±0.003 to 0.065± 0.004 mm made of high chemical purity thorium (99.9%) were manufactured in the RIAR (Dimitrovgrad, Russia). Each thorium foil of 17x50mm$^2$ was placed between aluminum and copper foils of the same size. Al and Cu were used to monitor the proton beam passing through the targets. The thickness of Al and Cu was 0.100 ± 0.004 and 0.038 ± 0.002 mm, respectively.

Irradiations of two foil stacks were performed at the linear accelerator of INR RAS [21] by protons with an initial energy of 158.5 and 100.1 MeV. The beam current was about 1 μA. Graphite degraders with a thickness of 2–4 mm were used to obtain the desirable proton energies. The calculation of energy and a current of protons bombarding the studied targets and monitors was performed using a modified STRAGL code [22], considering the straggling and scattering of protons. The cross section values of the monitor reactions $^{27}$Al(p, x)$^{22}$Na and $^{nat}$Cu(p, x)$^{62}$Zn recommended by the IAEA for the corresponding proton energy ranges [23] were used in the calculation: Al - 35-145 MeV; Cu - 15-100 MeV.

γ-Ray spectroscopy with high-pure Ge-detector ORTEC GEM15P4-70 was used for radionuclide determination. γ-Ray spectrometric measurements of Th foils were initiated approximately a day after the irradiation and were then carried out periodically. Spectra were analyzed using analyzing software Gamma Vision 32 and BNL-database [24]. The details of calculations can be found in our previous paper [25].

Theoretical calculations of cross sections were carried out utilizing the ALICE-IPPE nuclear code [26].

*Separation of the irradiation products*

The irradiated Th was dissolved in 7 M hydrochloric acid with the addition of HF up to $10^{-3}$-$10^{-4}$ M. The $^{230,233}$Pa fraction containing impurities of $^{95}$Nb and $^{103}$Ru was isolated by extraction chromatography with TEVA resin (Triskem Int., France) according to the procedure reported [27]. The obtained solution was separated into two parts; each part was evaporated to dryness several times with HNO$_3$ for removal of HF. The residue was dissolved in 7 M nitric or hydrochloric acid and used as a spike in extraction experiments.

*Extraction experiments*

The solvent extraction experiments were conducted following the well-known procedure [28]. The organic solution containing a mixture of an extractant (1-octanol, methyl isobutyl ketone (MIBK), aliquate 336), and a solvent (dodecane for 1-octanol and toluene for aliquate 336, 1:1 by volume) was pre-equilibrated with the aqueous stock solutions. For each extraction experiment, 2.0 mL of an aqueous solution containing $^{230,233}$Pa, $^{95}$Nb, $^{126}$Sb, and $^{103}$Ru tracers was mixed with an equal volume of the organic solution. The tubes were shaken for 20 min at 21±2 °C. According to the preliminary kinetic studies, this time is enough to reach the extraction equilibrium. Afterward, 1.0 mL aliquots were taken from both phases for the determination of the radionuclides by γ-ray spectroscopy.

The following extraction experiment series were conducted:

1. The pilot set of samples was prepared with varying times of shaking of the tubes (kinetic study). The aqua phase contented 7 M $HNO_3$ and the shaking time varied from 0.5 to 60 min.
2. The second set of samples was prepared with various nitric and hydrochloric acids concentrations (1–9 M) taking into account the content of the acid in the spike solution.
3. The third set was repeated with 0.4 M thorium (IV) nitrate or chloride in the aqua phase as a macro component.
4. The next experiments were carried out only with a solution of 1-octanol as an organic phase. In the fourth set, the content of the aqua phase was fixed (7 M $HNO_3$) while the concentration of 1-octanol in the organic phase varied from 3 to 100%.
5. The fifth set was prepared at the constant nitric or hydrochloric acids concentration of 7 M varying the HF concentration in the aqua phase in the range $10^{-5} - 1$ M.
6. In the last set the aqua phase included the fixed concentration of $HNO_3$ (7 M) and HF (0.01 or 0.001 M) and varying addition of boric acid or aluminum nitrate as a masking agent.

Back-extraction of Pa from octanol-dodecane solution (1:1) and MIBK was studied as well. Pa was previously extracted with the organic phase from 7 M $HNO_3$ (about 50 mL of each phase). The phases were separated completely. 2.0 mL of the organic solution was mixed with an equal volume of the aqueous solution. The tubes were shaken for 20 min at 21±2 °C. 1.0 mL aliquots were taken from both phases for the determination of the radionuclides by γ-ray spectroscopy. The experiments with solutions of hydrofluoric acid, nitric acid, and their mixture, hydrochloric, oxalic acid of different concentrations were carried out.

Measured activities were corrected for decay. Uncertainties of activity measurements have not exceeded 10 %.

**Results and discussion**

*The yield of $^{230}Pa/^{230}U$ in the reaction of natural thorium with medium-energy protons*

The cross sections of the $^{230}Pa$ formation obtained experimentally are shown in **Fig. 1** in comparison with the literature data [2,4,29-34] and the results of theoretical calculations and **Table 1**. $^{230}Pa$ is the product of one nuclear reaction: $^{232}Th$ (p,3n) $^{230}Pa$. Irradiations with different initial energies made it possible to determine cross sections with good accuracy and consistency with the literature data in a wide energy range of 35-140 MeV. The results of the theoretical calculation are noticeably overestimated in the studied energy range.

Other protactinium isotopes were also determined in γ-ray spectra of irradiated thorium targets: $^{228}$Pa ($T_{1/2}$ = 22.1 h), $^{229}$Pa ($T_{1/2}$ = 1.5 d), $^{232}$Pa ($T_{1/2}$ = 1.32 d), and $^{233}$Pa (27.0 d). The first three radionuclides are formed by reactions with neutron emission, and $^{233}$Pa produces mainly as a result of a reaction of Th with secondary neutrons: $^{232}$Th (n,γ) $^{233}$Th (22 min, β$^-$, 100%) → $^{233}$Pa. The number of secondary neutrons depends on the irradiation parameters (the initial energy of protons, the energy of protons entering and leaving the target) and the configuration of the irradiation chamber (the amount of cooling water and other materials exposed under irradiation).

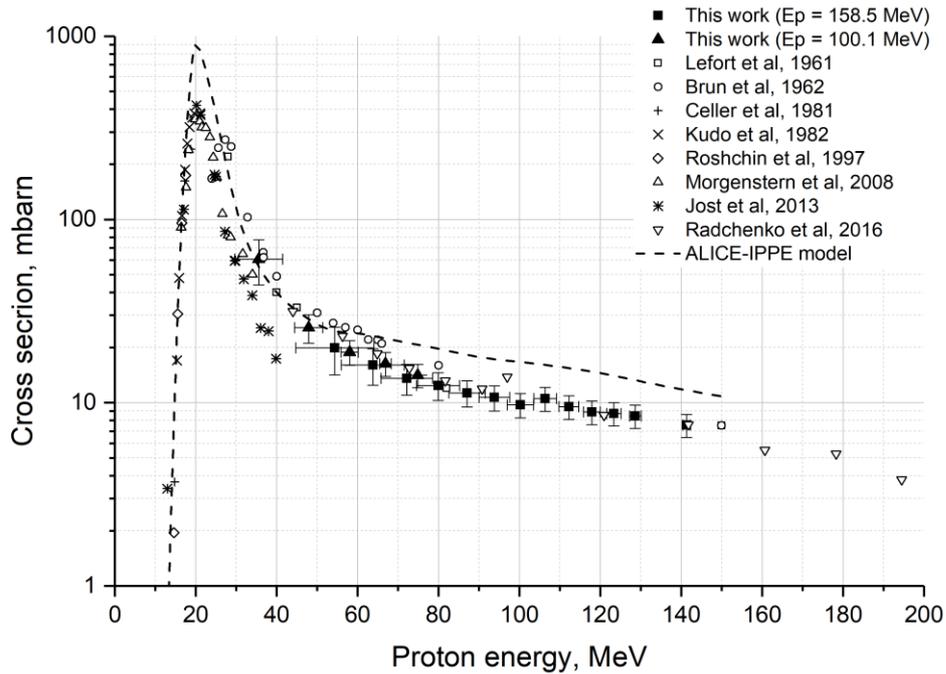

**Fig. 1** Experimental and theoretical (calculated by ALICE-IPPE model) $^{230}$Pa formation cross sections [2,4,29-34].

**Table 1.** Experimental cross sections for reaction $^{232}$Th (p,3n) $^{230}$Pa.

| Proton energy on target, MeV | Cross section, mbarn | Proton energy on target, MeV | Cross section, mbarn |
|---|---|---|---|
| Initial proton energy 158.5 MeV | | Initial proton energy 100.1 MeV | |
| 141.3 ± 1.0 | 7.5 ± 1.1 | 74.8 ± 0.9 | 14.1 ± 2.1 |
| 128.6 ± 1.5 | 8.5 ± 1.2 | 66.9 ± 1.4 | 16.4 ± 2.4 |
| 123.3 ± 1.8 | 8.7 ± 1.3 | 58.0 ± 2.2 | 18.9 ± 2.9 |
| 117.9 ± 2.1 | 8.9 ± 1.3 | 47.9 ± 3.5 | 25.7 ± 4.5 |
| 112.2 ± 2.4 | 9.5 ± 1.4 | 35.6 ± 5.9 | 60.7 ± 16.8 |
| 106.3 ± 2.8 | 10.5 ± 1.6 | | |
| 100.2 ± 3.3 | 9.7 ± 1.5 | | |
| 93.8 ± 3.8 | 10.7 ± 1.7 | | |
| 87.1 ± 4.5 | 11.3 ± 1.8 | | |
| 79.9 ± 5.3 | 12.4 ± 2.1 | | |
| 72.2 ± 6.4 | 13.6 ± 2.6 | | |
| 63.8 ± 7.8 | 16.1 ± 3.6 | | |
| 54.4 ± 9.7 | 20.0 ± 5.8 | | |

The formation of uranium isotopes during irradiation is going on through the β-decay of protactinium parent isotopes. Three isotopes of uranium are forming:

- $^{232}$Th (p,3n) → $^{230}$Pa (β$^-$, 7.8%) → $^{230}$U;

- $^{232}$Th (p,n) → $^{232}$Pa (β$^-$, 100%) → $^{232}$U (T$_{1/2}$ = 68.9 y);

- $^{232}$Th (n,γ) $^{233}$Th (β$^-$, 100%) → $^{233}$Pa (β$^-$, 100%) → $^{233}$U (T$_{1/2}$ = 1.6·10$^5$ y).

After 10 days of irradiation, the maximum amount of the goal $^{230}$U accumulates in 23 days. The calculation of the physical yield of $^{230}$U and long-lived impurities $^{232}$U and $^{233}$U in a thick target was carried out for this mode of irradiation and cooling. Yield values of these radionuclides depending on the energy of protons entering a thick thorium target, with fixed output energy equal to the threshold energy E$_{th}$ = 13.7 MeV of the formation of $^{230}$Pa are shown in **Fig. 2**. The yield of $^{230}$U was calculated using the experimental cross sections of $^{230}$Pa formation published in the work [2] for proton energies below 35 MeV, and the results of the present work (**Table 1**) for proton energies above 35 MeV. The $^{232}$U impurity was calculated using experimental cross sections of $^{232}$Pa formation from the paper [4], and the $^{233}$U impurity was calculated based on experimentally determined in the present work activities of $^{233}$Pa for the initial proton energy of 159 MeV.

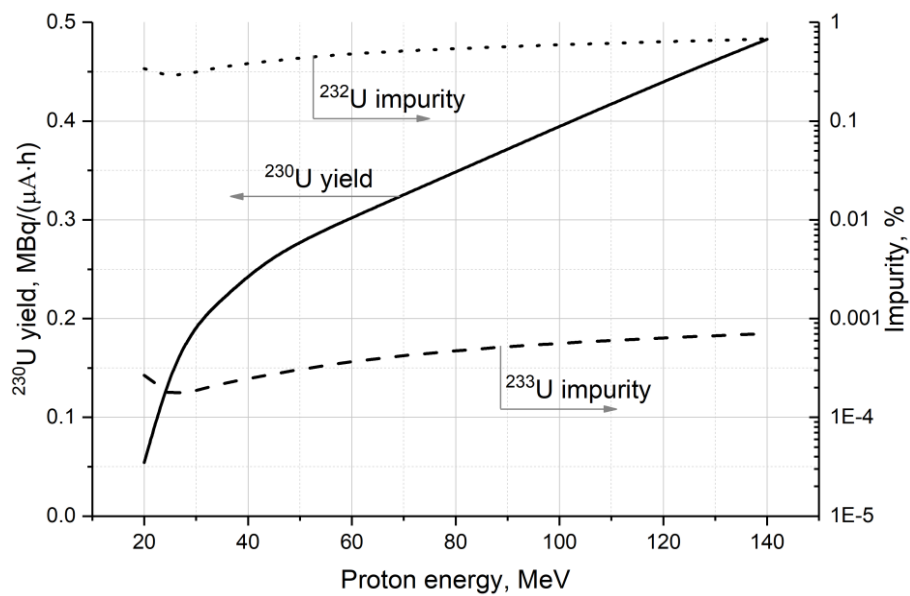

**Fig. 2** The yield of $^{230}$U, $^{232}$U, and $^{233}$U depending on the energy of protons bombarding a thick thorium target at fixed output energy of 13.7 MeV (calculation for 10-day irradiation and 23-days accumulation of $^{230}$U).

The yield of $^{230}$U reaches 0.48 MBq/(μA·h) at the initial energy of protons of 140 MeV. Protons with energy below 40 MeV make the greatest contribution to the formation of $^{230}$Pa/$^{230}$U. The range of proton energy of 60-140 MeV is optimal for the formation of $^{225}$Ac [15]. The yield of $^{230}$U in this energy range decreases to 0.18 MBq/(μA·h), which is comparable to the value of 0.24 MBq/(μA·h) obtained by Morgenstern et. al [2] for proton energies of 15-35 MeV.

The impurity of $^{233}$U is small and falls within $(2-7) \cdot 10^{-4}$%. $^{233}$U decays into a long-lived $^{229}$Th, locking up a further chain of decay. The admixture of $^{233}$U is decreasing significantly with a decrease in the initial energy of protons. The impurity of $^{232}$U reaches 0.7%, which is an obstacle to the direct medical application of $^{230}$U, however, $^{230}$U with an admixture of $^{232,233}$U can be used as a source for $^{230}$U/$^{226}$Th generator [35]. This case is similar to the pair of $^{225}$Ac/$^{213}$Bi with an impurity of $^{227}$Ac with the difference that the level of the impurity of $^{232}$U can be regulated by cooling time after irradiation and radiochemical separation of protactinium from uranium (**Fig. 3**).

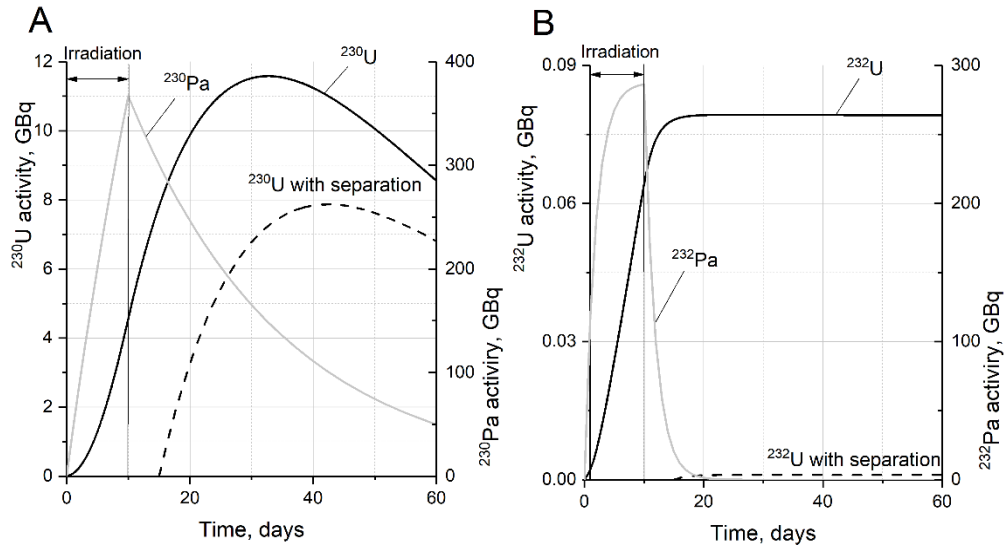

**Fig. 3** Accumulation of $^{230}$U (A) and $^{232}$U (B) for a thorium target irradiated for 10 days with a proton current of 100 μA and $E_p$ below 140 MeV. Dotted lines show the accumulation of $^{230}$U and $^{232}$U in the protactinium fraction separated from the uranium fraction 5 days after the EOB.

The impurity of $^{232}$U can be reduced by 50 times if the protactinium and uranium fractions are separated on the 5th day after the end of irradiation due to the shorter half-life of the parent

$^{232}$Pa (1.3 days versus 17.4 days of $^{230}$Pa). The yield of $^{230}$U decreases by about 1.5 times and the maximum accumulation is reached in 5 + 27 = 32 days after the end of bombardment while the admixture of $^{233}$U increases slightly.

The cooling of the target for 4-5 days and the subsequent accumulation of $^{230}$U in isolated Pa fraction for 27 days is in perfect consistency with the optimal scheme for producing the other medical radionuclide $^{225}$Ac [10,36]. Up to 3 GBq $^{230}$U with an admixture of 0.02% $^{232}$U and 0.001%, $^{233}$U can be obtained along with isolation of about 100 GBq $^{225}$Ac from the same thorium target. The activity of $^{230}$U is quite sufficient for the development of therapeutic drugs labeled with $^{230}$U or $^{226}$Th and conducting preclinical and clinical trials. The production of $^{230}$U for wide medical use can be organized via conventional compact cyclotrons that accelerate protons to low energies.

*Separation of Pa from irradiated thorium by liquid-liquid extraction*

The next task of our work is the development of a procedure that allows the simultaneous production of $^{225}$Ac and $^{230}$Pa/$^{230}$U from the same irradiated thorium target. Our goal is to propose a technologically adaptable approach to selective $^{230}$Pa isolation complying with the method of $^{225}$Ac production established previously [36]. The widely used extraction chromatography method has limitations, namely, relatively low radiation resistance of organic sorbents [37]. For a large volume of the initial solution, column separation becomes time-consuming and has reduced efficiency in the case of solutions containing a macro component. Therefore, we have chosen the liquid-liquid extraction method for Pa isolation. An important advantage of the extraction method is the ability to quickly isolate both large and ultra-small amounts of a substance.

Protactinium easily forms anionic complexes in acid solutions, therefore it can be extracted by oxygen-containing solvents capable of attaching protons in mineral acid solutions. Compounds consisting of an anionic complex of protactinium and an oxonium-type cation formed by the extractant are transferred into the organic phase. Since protactinium tends to form strong chloride complexes, hydrochloric acid solutions of Pa are more resistant to hydrolysis, and the extraction efficiency from hydrochloric acid solutions is usually higher than from nitric acid solutions [38].

The possibility for isolation of protactinium by LLE from a hydrochloric and nitric acids solution of irradiated thorium by various classes of organic extractants was analyzed. We considered tributyl phosphate (TBP), trioctylphosphine oxide (TOPO), di-2-ethylhexylphosphoric acid (HDEHP), methyl isobutyl ketone (MIBK), aliquate 336, and 1-octanol. These extractants and their analogs are most often used for the isolation and concentration of protactinium [39].

TBP, TOPO, and HDEHP make it possible to efficiently transfer Pa to the organic phase at high concentrations of hydrochloric or nitric acids in the aqueous phase, but do not provide selectivity concerning thorium [40,41]. The macro amounts of thorium follow Pa under the same conditions of extraction, which makes further separation problematic.

Preliminary experiments were carried out for (MIBK), aliquate 336, and 1-octanol on the isolation of protactinium at different concentrations of thorium in hydrochloric and nitric acid solutions. MIBK allows extracting Pa with high efficiency from hydrochloric acid solutions with a concentration of more than 5 M. 15% Ru, 20% Nb, and more than 95% Sb pass into the organic phase from 5 M HCl in addition to Pa (99% was extracted with MIBK, $V_{org}=V_{aq}$). The extractability of impurities grows with an increase in the acidity of the solution. The purification of Pa from these irradiation products causes certain difficulties because of the similarity of the behavior of Nb and Pa, as well as the complex and diverse chemistry of Ru [4]. The efficiency of Pa extraction from nitric acid solutions is low. The back extraction can be effectively provided by solutions of HF (≥0.1M) or oxalic acid (≥0.9M).

Aliquat 336 is extracted more than 98% protactinium with a 50% solution in toluene ($V_{org} = V_{aq}$) already at a concentration of hydrochloric or nitric acid of more than 3 M, while there was no noticeable difference in the efficiency of extraction in these acids. It was found that an increase in the concentration of thorium significantly reduces the extractability of protactinium. The extraction of protactinium drops to 64.8±5% for 7 M nitric acid and 48.5±5% for 7 M hydrochloric acid at a thorium concentration of 0.5 M. It restricts the applicability of this extractant for $^{230}$Pa isolation from irradiated thorium.

According to the literature data [42], aliphatic alcohols have a high selectivity of Pa extraction. There was found no significant difference in extraction of Pa from the alcohols chain length or position of OH- group so long as the extractant is water-immiscible [42]. We have been investigated extraction with 1-octanol which has already been tested in literature for Pa separation by LLE and extraction chromatography (sorbent TK400, Triskem Int.). The efficiency of Pa extraction with 1-octanol (1:1 in dodecane) depending on the concentration of the nitric and hydrochloric acid in the aqua phase is shown in **Fig. 4**.

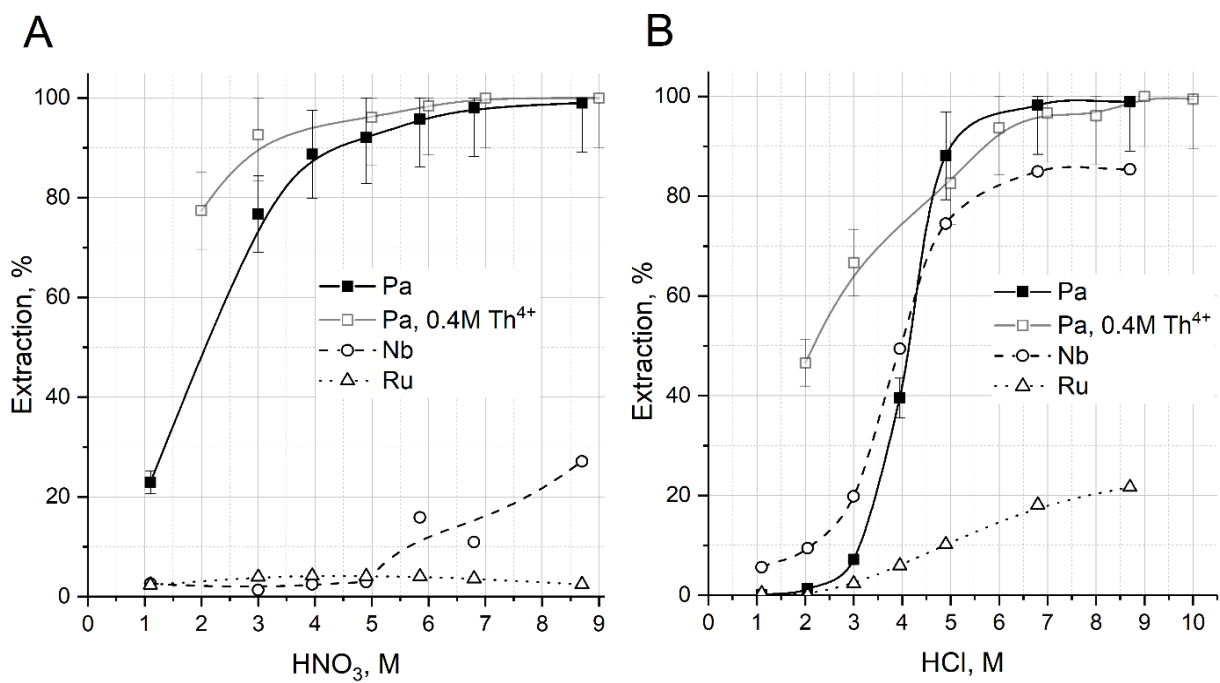

**Fig. 4** Extraction of Pa, Nb, and Ru with 1-octanol (1:1 in dodecane) depending on the acid concentration in the aqua phase: (A) for nitric acid solutions; (B) for hydrochloric acid solutions. The results of Pa extraction in the presence of 0.4 M $Th^{4+}$ in the aqua phase are shown with the gray curve.

It can be seen that the extraction efficiency of protactinium exceeds 90% for both mineral acids at their concentration in the aqueous phase of more than 5 M. These results are in good agreement with the literature data for 1-octanol [42], 1-ethylhexanol [42], 2,4-dimethyl-3-heptanol [43] and 2,6-dimethyl-4-heptanol [42,44]. The presence of macro amounts of thorium (0.4 M) does not reduce the efficiency of Pa extraction in the strongly acidic region (**Fig. 4**). Extraction is noticeably higher in the presence of thorium in the area of low acid concentration. Thorium nitrate/chloride molecule dissociates to form 4 anions, therefore, the concentration of nitrate and chloride ions in the presence of 0.4 M thorium is up to 1.6 M. A higher concentration of the counterion increases the extraction efficiency, and the lower the acid concentration, the more noticeable the effect. Thus, it was shown [42] that the extraction dependence of Pa for 2,6-dimethyl-4-heptanol on the counter-ion concentration in logD-log[$An^-$] coordinates has a slope of 5.5-5.7 for 1-4 M HCl and 1.9-2.1 for 1-4 M $HNO_3$.

The dependence of the extraction of two other products of thorium irradiation, Nb and Ru, on the acid concentration is also given in **Fig. 4**. It can be seen that the efficiency of their extraction from nitric acid solutions is noticeably inferior to hydrochloric acid solutions, less than 15% Nb and 5% Ru pass from 7 M $HNO_3$ into the organic phase. Sb remains completely in the aqueous phase under these conditions. We can conclude that nitric acid solutions make it possible to obtain a purer fraction of $^{230}$Pa. This correlated with the fact that the $^{225}$Ac isolation procedure involves dissolving the target in nitric acid [36], therefore, only nitric acid solutions were further investigated as an aqua phase for LLE of Pa.

A study of the kinetics of Pa extraction with 1-octanol solution in dodecane (1:1) demonstrated that the extraction process is fast and the equilibrium in the system is established in less than 1 min.

The dependence of the extraction efficiency of Pa, Nb, and Ru on the concentration of the extractant in the organic phase is shown in **Fig. 5**. The slope analysis in logD-log[1-octanol] coordinates allows us to determine the apparent stoichiometry of the process. The slope is 1.51 ± 0.06, which is in good agreement with the literature data. Namely, Knight reported that for extracting Pa from 6 M HCl (other conditions are similar) the slope is 1.58±0.08 [42]. Therefore, we can suggest that protactinium is extracted as a monobasic complex of general formula HPa(OH)$_m$(NO$_3$)$_n$ · 1–2(1-octanol), with m + n = 6, and n ≥ 1.

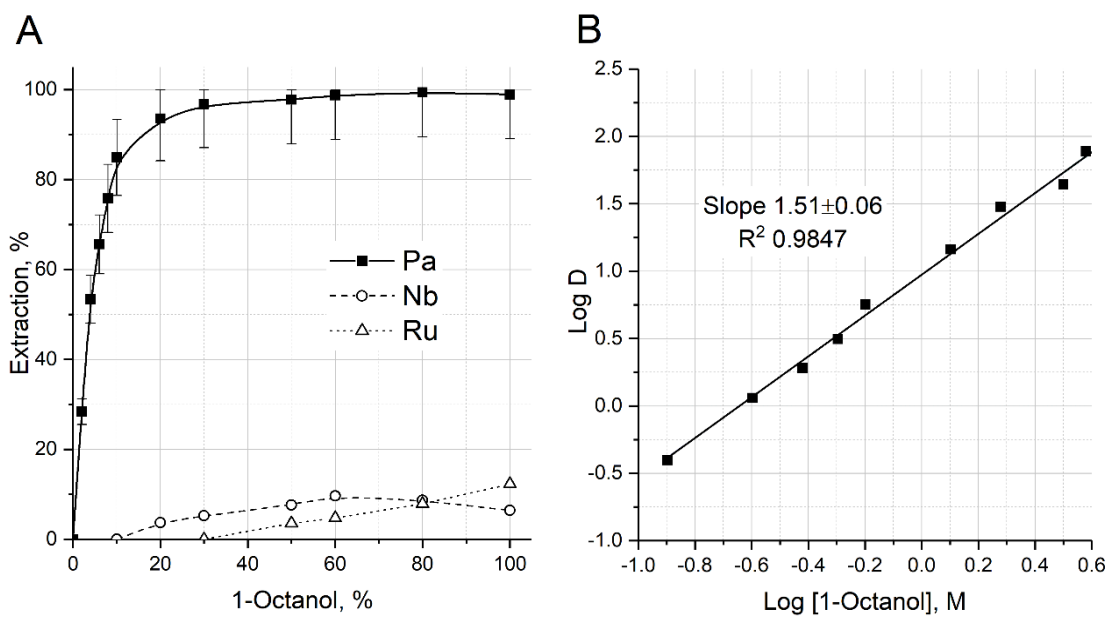

**Fig. 5** (A) Extraction of Pa, Nb, and Ru with 1-octanol (in dodecane) from 7 M HNO$_3$ depending on the 1-octanol concentration in the organic phase; (B) Plot of distribution ratios of Pa versus 1-octanol concentration in the organic phase.

It is well known that Pa forms strong complexes with fluoride and oxalate ions. This property is often used in separation chemistry, for protactinium leaching and back-extraction. In particular, the not-extractable PaF$_7^{2-}$ complex is formed with fluoride ions [39]. Dissolving the irradiated thorium in hydrochloric or nitric acids requires the addition of HF, which is necessary to destroy the thorium oxide film formed on the metal surface. The developed technology of $^{225}$Ac production includes the addition of HF up to $10^{-2}$ M for Th dissolution [36]. Preliminary experiments have shown that this concentration may reduce the efficiency of the subsequent isolation of Pa by LLE.

The effect of the addition of HF on extraction with a solution of 1-octanol in dodecane from 7M HCl and HNO$_3$ was investigated (Fig. 6). It can be seen that hydrofluoric acid up to a concentration of $10^{-4}$ M or $10^{-3}$ M may be added to 7 M nitric or hydrochloric acid solutions of protactinium, respectively, without greatly lowering the extraction. At higher concentrations of fluoride, the extraction falls rapidly to zero. Chloride complexes of Pa are stronger than nitrate ones, therefore the influence of fluoride ions on nitric acid solutions is more pronounced.

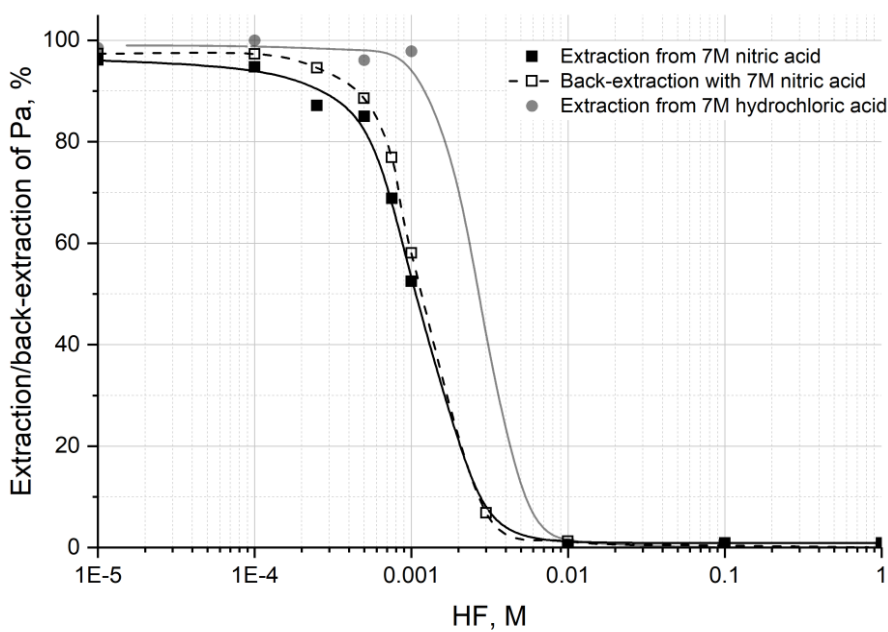

Fig. 6 Extraction of Pa with 1-octanol (1:1 in dodecane) from 7 M nitric and 7 M hydrochloric acids depending on the HF concentration in the aqua phase (the solid lines). Back-extraction of Pa with from 1-octanol (1:1 in dodecane) depending on the HF concentration in 7 M HNO$_3$ is shown by the dashed line.

Masking agents can be added to restore the extractability of Pa into the organic phase in the presence of HF. Aluminum salts and boric acid are most often used in these cases. They form fluoride complexes competing for F$^-$ ions and destroying fluoride complexes of Pa [39,45]. **Figure 7** shows the extraction of Pa with a solution of 1-octanol in dodecane from 7 M HNO$_3$ with the addition of HF, depending on the concentration of the masking agents in the aqueous phase. The determined experimentally solubility of boric acid in 7 M HNO$_3$ does not exceed 0.3 M. It can be seen that the aluminium salt is more effective for binding fluoride ions than boric acid at the same concentration. Thus, at least 0.25 M $Al^{3+}$ is needed to restore the extractability of Pa in the presence of 0.01 M HF, while boric acid is less effective.

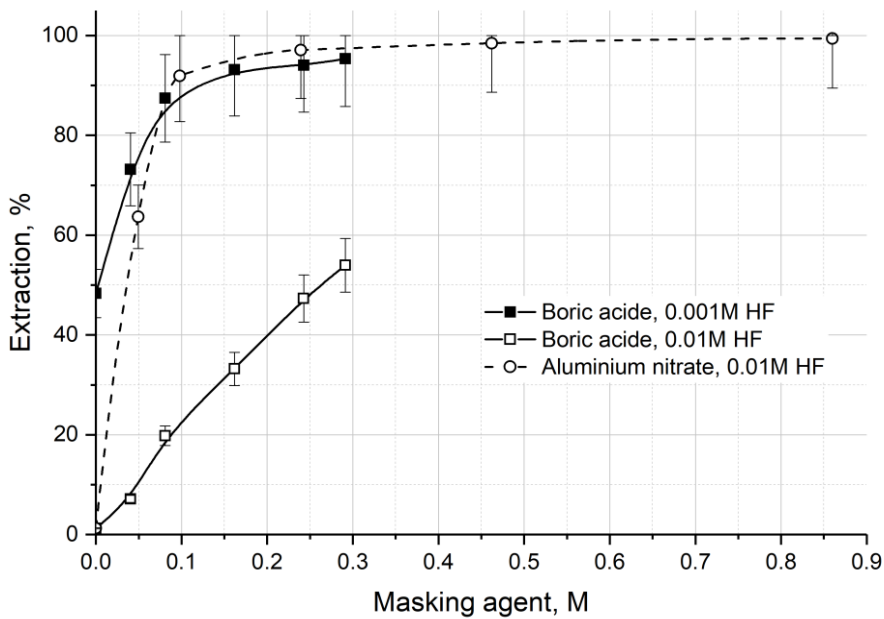

**Fig. 7** Extraction of Pa with 1-octanol (1:1 in dodecane) from 7 M nitric acid depending on the concentration of masking agents in the aqua phase at fixed content of HF.

Solutions containing fluoride ions are appropriate for Pa back-extraction. The dependence of Pa back-extraction from a solution of 1-octanol in dodecane on the content of $F^-$ in 7 M $HNO_3$ is represented by the dashed line in **Fig. 6**. It can be seen that the extraction and back-extraction curves are almost identical, which indicates that the process is equilibrial and reversible. The addition of 0.01 M HF is sufficient for quantitative back-extraction of Pa into the aqueous phase. The back-extraction with oxalic acid solutions is also effective and allows Pa to be quantitatively transferred to the aqueous phase, however, further removal of oxalic acid may be problematic. Dilute solutions of hydrochloric or nitric acids do not allow extraction of more than 80% of Pa ($V_{org} = V_{aq}$). Pa in such solutions is extremely unstable tends to hydrolyze and adsorb on the surface of glassware [16].

Thus, the $^{230}$Pa separation procedure includes the following steps. The irradiated thorium is dissolved in 7-8 M nitric acid with the addition of HF, and corresponding amounts of a masking agent should be added to bind free fluoride ions. Then Pa is extracted with a solution of 1-octanol in dodecane (1:1), pre-equilibrated with 7-8 M $HNO_3$, while the aqueous phase is further processed according to the established route for $^{225}$Ac production [10,36]. Finally, Pa is back-extracted with a small volume of acid with the addition of 0.01 M HF. $^{230}$Pa is an intermediate product for nuclear medicine, so the back-extracted protactinium stabilized as a fluoride complex is further stored for $^{230}$U accumulation. The $^{230}$U can be easily separated chromatographically, for example, on TEVA resin [27]. If additional purification is required, a masking agent (0.25 M $Al^{3+}$) is added to the resulting solution again and Pa can be purified using ion exchange [4,17] or extraction chromatography [4,6,18,19]).

The proposed procedure makes it possible to isolate and concentrate $^{230}$Pa from the irradiated thorium target with a yield of at least 97%. The main impurities in the isolated $^{230}$Pa are $^{95}$Nb (SF 6-10) and $^{103}$Ru (SF 20-25).

## Conclusion

The cross sections of the $^{230}$Pa formation in the energy range of 140-35 MeV were determined using the stacked-foil technique. The yields of the goal irradiation product $^{230}$U, as well as the main isotopic impurities $^{232}$U and $^{233}$U in a thick target irradiated by medium energy protons, have been estimated based on the obtained values and literature data. The yield of $^{230}$U is 0.18 MBq/(μA·h) at the energy of protons in the range of 60-140 MeV which is comparable to the yield for $^{230}$U producing on 30 MeV cyclotrons. The method of liquid-liquid extraction with 1-octanol provides quick and selective separation of Pa from the solution of the irradiated thorium. It is optimal to carry out the isolation of Pa with 5 days delay after the end of bombardment, and then store the fraction of $^{230}$Pa to accumulate $^{230}$U for 27 days. The developed procedure makes it possible to produce several of GBq of $^{230}$U per irradiation session as a by-product of $^{225}$Ac. The $^{230}$Pa separation technique can be as well applied for thorium irradiated on a cyclotron with proton energy $E_{max}$ of 30 MeV when $^{230}$U is the main product of irradiation.

Future work is supposed to be focused on upscaling the developed procedure to obtain high activity of $^{230}$U and loading the $^{230}$U/$^{226}$Th generator. Another important issue for investigation is the possible effect of masking agents ($Al^{3+}$, $H_3BO_3$) on the further extraction-chromatographic isolation of $^{225}$Ac (DGA resin sorbent, Triskem Int.).

## Acknowledgments

This work has been supported by the Russian Foundation for Basic Research under Contract № 20-53-15007.